\title{HELICITY MODULUS \\ AND \\ EFFECTIVE HOPPING \\
IN THE \\ TWO-DIMENSIONAL HUBBARD MODEL \\
USING SLAVE-BOSON METHODS.}
\author{P.J.H. Denteneer and M. Blaauboer\\
        Instituut-Lorentz, University of Leiden,\\
        P. O. Box 9506, 2300 RA Leiden, The Netherlands}
\date{}
\documentstyle[12pt]{article}
\newcommand{\be}{\begin{equation}}
\newcommand{\ee}{\end{equation}}
\newcommand{\bc}{\begin{center}}
\newcommand{\ec}{\end{center}}
\newcommand{\ha}{{\scriptstyle \frac{1}{2}}}
\newcommand{\eps}{\varepsilon}
\newcommand{\ffi}{\varphi}

\newcommand{\bk}{{\bf k}}

\newcommand{\bq}{{\bf q}}
\newcommand{\bQ}{{\bf Q}}
\newcommand{\eb}{\bar{\eps}}
\newcommand{\mub}{\bar{\mu}}
\newcommand{\lab}{\bar{\lambda}}
\newcommand{\mut}{\tilde{\mu}}
\newcommand{\lat}{\tilde{\lambda}}
\newcommand{\Ne}{{\cal N}(\eps)}
\newcommand{\Nve}{{\cal N}_{\rm v}(\eps)}
\newcommand{\pri}{\raisebox{.8ex}{$\prime$}}
\begin{document}
\hoffset=-0.0 true cm
\voffset=-2.0 true cm
\maketitle
\vspace{1cm}
\begin{abstract}
The slave-boson mean-field method is used to study the two-dimensional
Hubbard model. A magnetic phase diagram allowing for paramagnetism,
weak- and strong ferromagnetism and antiferromagnetism is constructed and
compared to the corresponding phase diagram using the Hartree-Fock
approximation
(HFA). Magnetically ordered regions are reduced by a factor of awi
bout 3 along
both the $t/U$ and density axes compared to the HFA. Using the spin-rotation
invariant formulation of the slave-boson method the helicity modulus is
computed
and for half-filling is found to practically coincide with that found using
variational Monte Carlo calculations using the Gutzwiller wave function.
Off half-filling the results can be used to compare with Quantum Monte Carlo
calculations of the effective hopping parameter. Contrary to the case of
half-filling, the slave-boson approach is seen to greatly improve the results
of the HFA when off half-filling.

\end{abstract}
\vspace{1.0cm}
Short title: Helicity modulus in the Hubbard model using slave bosons
\newpage

\section{Introduction}
In the study of correlated electrons, for which both charge and spin degrees
of freedom are relevant, the Hubbard model is an intriguing simplification
of reality that still contains a great deal of the essential physics
\cite{Hub}.
Although superconductivity has not been demonstrated in this model, it is able
to explain or reproduce a large number of experimental results on the
copper oxides which superconduct at high temperatures \cite{Dag}.
Despite such encouraging results, understanding of the model in dimensions
two and higher is still rather limited. Even when the most elementary
mean-field
approximation, the Hartree-Fock approximation, is invoked, the phase diagram
cannot be determined in full \cite{InLi}, since inhomogeneous phases, like
spiral phases or domain walls, are able to
supersede simple ferro- or antiferromagnetic phases.
The possibility exists that more complicated phases not considered sofar
are also important. Rigorous techniques like Quantum Monte Carlo and exact
diagonalization are limited to temperatures which may be too high and lattices
which may be too small, respectively.
Therefore it is of interest to employ approximations which go beyond the
Hartree-Fock approximation (HFA).

Some years ago Kotliar and Ruckenstein introduced,
for the Hubbard model,
the technique of using slave bosons to keep closer track of the site-occupancy
than is done in the HFA. If a further approximation is made, the so-called
``slave-boson mean-field'' (SBMF) approximation, this approach was shown to be
equivalent to the approximation scheme of Gutzwiller for the Hubbard
model \cite{KR}.
In a parallel development it was shown by Vollhardt and co-workers that the
Gutzwiller approximation scheme becomes exact in the limit of an infinite
number of
spatial dimensions, whereas the HFA does not become exact in this limit
\cite{MV,Voll}.
Therefore, just like in classical statistical physics mean-field theory is a
good starting point to study a specific model (because fluctuations become
increasingly less relevant when increasing the dimension), for quantum models
the SBMF approximation is a good starting point. In any case, it is expected
to be an improvement over the HFA. In this connection, it is of interest to
note
that Ole\'{s} and Zaanen compared the Gutzwiller approximation (GA) with
the HFA
in a two-band model of copper-oxide planes and showed
that the GA is a good approximation for correlations at small length scales
\cite{OlZa}.
More recently the slave-boson approach of Kotliar and Ruckenstein has been
refined in order to make it spin-rotation invariant \cite{LWH,FWII}.

In this paper, we make a detailed comparison of the SBMF and Hartree-Fock
approximations to the (one-band) Hubbard model on a two-dimensional
square lattice.
Where possible we also compare to Quantum Monte Carlo calculations.
By using the analytical result for the density of states of freely hopping
electrons on the square lattice all calculations can conveniently be
performed for a lattice of infinite size using one-dimensional integrals
(over energy) only. Although we will mostly present results for the
ground state ($T=0$), the formulation we present is to a large extent
valid for finite temperatures as well. A further advantage of the SBMF approach
is that, in principle, it is valid for the whole range of Hubbard repulsion
strengths and electron densities.
First, we construct a ground-state phase-diagram allowing only for the
simple magnetic phases: paramagnetic, antiferromagnetic, weakly and strongly
(i.e. partially and fully polarized) ferromagnetic.
We determine all first-order and continuous transitions between
these phases.\footnote{Some of the
results on the SBMF phase diagram were also obtained in Ref.\cite{Evans}.}
Although this approach may
not give much information on the exact phase diagram (for instance, it is
known that phases with spiraling magnetization supersede the antiferromagnet
when going off half-filling), it gives a clear impression of the improvement
of the SBMF approximation over the HFA. Next, we derive an expression for
the helicity modulus or spin stiffness in SBMF approximation
using the spin-rotation invariant formulation and compare
(for half-filling) to previous calculations of the helicity modulus in the
HFA, as well as using the Gutzwiller wave function in a variational Monte Carlo
calculation \cite{PDrhos}. It turns out that
in SBMF approximation, apart from a negligible contribution,
the helicity modulus is completely determined by the average kinetic energy
(as is the case in the HFA) and therefore equivalent to the effective
hopping parameter. When comparing the SBMF results for the effective hopping
parameter to HFA and QMC calculations
the improvement with respect to the HFA is only a few percent at half-filling
and practically coincides with the results of variational Monte Carlo
calculations using the Gutzwiller wave function, but the improvement is
substantial off half-filling.

The paper is organised as follows: in Section 2, we briefly introduce the
slave-boson mean-field method for the Hubbard model
and present the free energy for the antiferromagnetic,
ferromagnetic and spiral phases as well as the corresponding consistency
equations. In Section 3, we construct the ground-state phase diagram
for the simple
magnetic phases listed above and compare to the corresponding HFA phase
diagram.
An expression for the helicity modulus is derived and its connection to
the effective
hopping parameter is discussed in Section 4. The SBMF results are compared
to both
the HFA and QMC calculations where possible.
The last section contains a discussion
of the results and draws some conclusions.

\section{Slave Boson Mean Field method}
The Hamiltonian for the Hubbard model is given by: \\
\begin{equation}
{\cal H} = - \sum_{ i j \sigma } t_{ij} c_{i \sigma}^{\dagger}
c_{j \sigma}
+ U \sum_{i}n_{i \uparrow}n_{i \downarrow}-\mu \sum_{i \sigma}
n_{i \sigma},   \label{eq:HH}
\end{equation}
where $c_{i \sigma}^{\dagger}$ creates an electron at
site $i$ with spin $\sigma$,
$n_{i \sigma}=c_{i \sigma}^{\dagger}c_{i \sigma}$,
$t_{ij}$ is the one-electron transfer integral between sites
$j$ and $i$ ($t_{ij}$
equals $t$ if $i$ and $j$ are nearest neighbours and 0 otherwise),
$U$ the on-site repulsion ($U > 0$),
and $\mu$ the chemical potential ($\mu = U/2$ corresponds to a
half-filled lattice, i.e. $n = \sum_{i \sigma} \langle n_{i \sigma}
\rangle = 1$).
The Hubbard model allows for four different occupancies of a single site:
it can be empty, singly occupied by either a spin-up or spin-down electron,
or doubly occupied (by electrons of opposite spin).
This leads to the idea of introducing four different
kinds of ``slave'' bosons, one for each of the possible occupancies.
In order for this to become a bookkeeping device one introduces the
constraints that at each site there is always exactly one boson present and
that it is of the kind corresponding to the electron occupancy. If
$e$, $p_{\uparrow}$, $p_{\downarrow}$ and $d$ denote the annihilation operators
for the four kinds of bosons, these constraints are:
\begin{eqnarray}
& e^{\dagger}_i e_i + \sum_{\sigma} p^{\dagger}_{i\sigma} p_{i\sigma} +
d^{\dagger}_i d_i = 1  &~~~~\mbox{for all $i$} , \label{eq:Q1} \\
& c^{\dagger}_{i\sigma} c_{i\sigma} = p^{\dagger}_{i\sigma} p_{i\sigma} +
d^{\dagger}_i d_i &~~~~\mbox{for all $i$ and $\sigma = \uparrow,\downarrow$} ,
\label{eq:Q2}
\end{eqnarray}
The interaction term in the Hamiltonian (\ref{eq:HH}) can now be replaced by
one containing only the counting operator for $d$-bosons. In order for the
boson presence to keep in correspondence with the electron occupancy
for each site, the hopping
term in the Hamiltonian needs to be adjusted by connecting to each electron
annihilation operator $c_{i\sigma}$ the following {\em boson-transformation}
operator:
\be
\tilde{z}_{i\sigma} = e^{\dagger}_i p_{i\sigma} + p^{\dagger}_{i,-\sigma} d_i .
\label{eq:zt}
\ee
In fact this choice for $\tilde{z}$ is not unique \cite{Lav}
and we follow Ref.\cite{KR}
by making a choice which in the case of $U=0$ gives the correct result
if a subsequent saddle-point approximation is made:
\be
z_{i\sigma} = \left( 1 - d^{\dagger}_i d_i - p^{\dagger}_{i\sigma} p_{i\sigma}
\right)^{-\ha} \tilde{z}_{i\sigma} \left( 1 - e^{\dagger}_i e_i -
p^{\dagger}_{i,-\sigma} p_{i,-\sigma} \right)^{-\ha} . \label{eq:zz}
\ee
In the physical subspace, defined by (\ref{eq:Q1})-(\ref{eq:Q2}),
of the enlarged boson-electron Hilbert space, the Hamiltonian,
\be
{\cal H}_{\rm SB} =  - \sum_{ i j \sigma } t_{ij} c_{i \sigma}^{\dagger}
z_{i \sigma}^{\dagger}z_{j \sigma}c_{j \sigma}
+ U \sum_{i} d^{\dagger}_i d_i - \mu \sum_{i \sigma} n_{i \sigma},
  \label{eq:HSB}
\end{equation}
has the same matrix elements as the original Hamiltonian in the original
Hilbert space containing only electron states.
Therefore, up to here only a reformulation of the original problem has been
achieved. However, if now in the functional integral formulation
the saddle-point
approximation of time- and position-independent Bose fields is made, a set of
equations results
which is similar to those found in the Hartree-Fock approximation,
but more general. This approximation is called the Slave-Boson Mean-Field
(SBMF)
approximation. For comparison,
the HFA can be obtained in such a functional integral formulation
by first applying a suitable Hubbard-Stratonovich transformation and then
making
the saddle-point approximation for the original problem \cite{Fradkin}.
For further details on the functional integral formulation for the Hubbard
model and the subsequent saddle-point approximation we refer to previous papers
on this subject \cite{KR,LWH,FWII,Lav,FWI}.

In the next three subsections, we present the expressions for the free
energy of the ferromagnetic, antiferromagnetic, and spiral phases. For the
ferromagnet and antiferromagnet we also give the consistency equations
for the appearing self-consistent fields, which must be obeyed for the
free energy
to become minimal. To some extent this is a repetition of previously
published results \cite{Evans,FWI}, but they are given for the reader's
convenience and to establish the notation.

\subsection{Ferromagnetic Phase}
For the ferromagnetic phase one assumes a non-zero homogeneous magnetization
$m$,
even if a magnetic field is absent.
To be able to calculate the magnetic susceptibility we also include a magnetic
field $h$ in the Hamiltonians ${\cal H}$ and ${\cal H}_{\rm SB}$ by
adding a term:
\be
{\cal H}_{\rm mag} = - h \sum_{i \sigma} \, \sigma n_{i \sigma} ~.
\label{eq:Hmag}
\ee
Here and in the following we adopt the convention that if $\sigma$ does
not appear
as an index it attains the values $+1$ and $-1$ if the corresponding index is
$\uparrow$ and $\downarrow$, respectively.
In order to treat the density of electrons $n$ and magnetization $m$
on an equal footing,
we introduce a slightly different definition for the free energy per site, to
be denoted by $\ffi$. For the ferromagnet our free energy is defined as:
\be
\ffi_{\rm  F} = - \frac{1}{\beta N} \ln \mbox{Tr} e^{-\beta ({\cal H}_{\rm SB}
+
{\cal H}_{\rm mag})} + \mu n + hm ~. \label{eq:fFdef}
\ee
It is given by (see also Refs.\cite{KR}, \cite{Evans}):
\be
\ffi_{\rm  F} = - \frac{1}{\beta N} \sum_{\bk , \sigma}\, \ln \left[ 1 +
\mbox{e}^{-\beta E_\sigma(\bk)} \right] + Ud^2 + \mub n + \lab m ~,
\label{eq:fF}
\ee
where
\be
E_\sigma(\bk) = q_\sigma t(\bk) - \sigma \lab - \mub ~, \label{eq:Eksig}
\ee
with
\be
t(\bk) = -2t(\cos k_x + \cos k_y) ~, \label{eq:tk}
\ee
the band structure of freely hopping
electrons on a square lattice. The density $n$ and magnetization $m$
are given by: $n = n_{\uparrow} + n_{\downarrow}$ and
$m = n_{\uparrow} - n_{\downarrow}$ with $n_{\sigma} =
\langle n_{i \sigma} \rangle $.
Since in principle the electrons have been integrated out in obtaining
(\ref{eq:fF}),
$n$ and $m$ can be understood as a shorthand for the Bose fields $p_{\uparrow}$
and $p_{\downarrow}$ via the relation: $n_\sigma = p_\sigma^2 + d^2$ (which is
the average of constraint (\ref{eq:Q2})). The parameters $\mub$ and $\lab$
are an effective chemical potential and effective magnetic field, respectively,
which incorporate the Lagrange multipliers $\lambda_\sigma^{(2)}$ used to
enforce the constraints (\ref{eq:Q2}):
\begin{eqnarray}
\mub & = & \mu - \ha \left( \lambda_{\uparrow}^{(2)} +
\lambda_{\downarrow}^{(2)} \right) ~, \\ \label{eq:mub}
\lab & = & h - \ha \left( \lambda_{\uparrow}^{(2)} -
\lambda_{\downarrow}^{(2)} \right) ~. \label{eq:lab}
\end{eqnarray}
The Lagrange multiplier $\lambda^{(1)}$ associated with constraint
(\ref{eq:Q1})
has disappeared because the average of this constraint must hold. In fact, both
constraints are only satisfied on average in the saddle-point approximation.
The band-renormalization factor $q_\sigma$ appearing in (\ref{eq:Eksig}) is
in this approximation of time- and position-independent Bose fields
a function of $n, m$ and $d$ (as follows directly from
(\ref{eq:zt})-(\ref{eq:zz})):
\be
q_\sigma (n,m,d) \equiv \langle z^{\dagger}_{j\sigma}z_{i\sigma} \rangle =
\frac{\left[ \sqrt{ (1-n+d^2)(n+\sigma m -2d^2)} + d \sqrt{n-\sigma m -2d^2}
\right]^2}{ (n+\sigma m) \left[ 1 - \ha(n + \sigma m) \right] } ~.
\label{eq:qsig}
\ee
We note that if this $q_\sigma$ is rewritten as a function of
$n_{\uparrow}, n_{\downarrow}$ and $d$ one exactly recovers the expression
for the band renormalization in the Gutzwiller approximation
(see Ref.\cite{VollRMP};
our $d^2$ is called $d$ there\footnote{In Ref.\cite{OlZa} similar
renormalization factors are derived in an alternative manner.}).
This would not be true if another choice for
$z_{i\sigma}$ than (\ref{eq:zz}) had been made.

The sum over $\bk$ in (\ref{eq:fF}) is over the whole Brillouin zone of the
square lattice. In the limit of an infinitely large lattice, which can be
treated after the approximations made, the resulting integral over the
(two-dimensional) Brillouin zone can be rewritten as an (one-dimensional)
integral over energy using the density of states (DOS) of freely
hopping electrons.
For the square lattice this DOS, $\Ne$, is known analytically:
\be
{\cal N}(\eps) \equiv \frac{1}{N} \sum_{\bk} \, \delta( \eps - t(\bk) ) =
\left\{ \begin{array}{cl}
\frac{1}{2\pi^2 t} K \left[ 1 - \left( \frac{\eps}{4t} \right)^2 \right] &
|\eps| \leq 4t \\
0 & |\eps| > 4t
\end{array} \right. ~, \label{eq:DOS}
\ee
where $K(x)$ is the complete elliptic integral of the first kind \cite{AS}.
Employing the DOS the free energy is:
\be
\ffi_{\rm F} = - \frac{1}{\beta} \sum_{\sigma}
\int \mbox{d}\eps \, \Ne \ln \left[ 1 +
\mbox{e}^{-\beta E_\sigma(\eps)} \right] + Ud^2 + \mub n + \lab m ~,
\label{eq:fFN}
\ee
with $E_\sigma(\eps) = q_\sigma \eps - \sigma \lab - \mub$.
Unless stated otherwise explicitly integrals over $\eps$ run from $-\infty$ to
$\infty$ (the relevant integration range is of course limited by
(\ref{eq:DOS})).

For a given interaction strength $U$, $\ffi_{\rm F}$ is now given as a
function of the
five variables $n, m, d, \mub$ and $\lab$, whereas $\mu$ and $h$ are control
parameters, regulating (although not directly in this slave-boson approach) the
density and magnetization. The optimal values for the five variables must be
found from the three minimization conditions:
$\partial \ffi_{\rm F}/\partial d = 0$,
$\partial \ffi_{\rm F}/\partial \mub = 0$,
$\partial \ffi_{\rm F}/\partial \lab = 0$, as well as from
the two equations arising from the Legendre transform between grand potential
(function of $\mu$ and $h$) and free energy (function of $n$ and $m$):
$\partial \ffi_{\rm F}/\partial m = h$ and
$\partial \ffi_{\rm F}/\partial n = \mu$. Applying these
conditions to (\ref{eq:fFN}) results in:
\begin{eqnarray}
U & = & - \frac{1}{2d} \sum_\sigma \, q_{\sigma d} \eb_\sigma ~,
\label{eq:ceUF}\\
n & = & \sum_\sigma  \, n_\sigma ~, \label{eq:cenF} \\
m & = & \sum_\sigma  \, \sigma n_\sigma ~, \label{eq:cemF} \\
\lab & = & h - \sum_\sigma \, q_{\sigma m} \eb_\sigma ~,\label{eq:celaF} \\
\mub & = & \mu - \sum_\sigma \, q_{\sigma n} \eb_\sigma ~, \label{eq:cemuF}
\end{eqnarray}
where $q_{\sigma \alpha}$ denotes the first partial derivative of
$q_\sigma$ with
respect to $\alpha$ (= $n$, $m$, $d$) and we introduced the abbreviations:
\begin{eqnarray}
n_\sigma & = & \int \mbox{d}\eps \, \Ne f \left[ E_\sigma(\eps) \right] ~, \\
\eb_\sigma & = & \int \mbox{d}\eps \, \Ne \eps f
\left[ E_\sigma(\eps) \right] ~,
\end{eqnarray}
with the Fermi-Dirac distribution,
$f(E) = \left[ 1 + \mbox{e}^{\beta E} \right]^{-1}$.
The five equations (\ref{eq:ceUF})--(\ref{eq:cemuF})
need to be solved self-consistently
for given $U$, $\mu$ and $h$ and the results can be inserted in
(\ref{eq:fFN}) to
obtain the corresponding free energy. In practice, we will often be interested
in calculations for a fixed density $n$, in which case the last equation
(\ref{eq:cemuF}) does not appear ($\mu$ disappears from the problem,
however $\mub$ remains).
Expressions for partial derivatives of $q_\sigma$ are given in Appendix A.

The above allows to compute the free energy of three different phases:
the paramagnetic phase (PM), for which $m=0$ if $h=0$,
the strong ferromagnet (SF), for which
$m=n$ and both are non-zero even if $h=0$, and the weak ferromagnet (WF),
for which $m<n$ and both are non-zero even if $h=0$. In Section 3, we
will obtain
the lines in the ($t/U,n$)-diagram
of first-order phase transitions between these phases. In order to obtain the
line of the continuous phase transition between paramagnet and ferromagnet
(the ferromagnet can be either strong or weak),
one needs to find where the susceptibility $\chi$
of the paramagnet diverges. An expression for $\chi$ is derived in Appendix B.

\subsection{Anti-ferromagnetic Phase}
For the antiferromagnetic phase we divide the square lattice in two
sublattices,
such that points on one sublattice have only points of the other sublattice as
nearest neighbours. Furthermore, we
assume a non-zero staggered magnetization $m_{\rm s}$,
i.e. the magnetization is $m_{\rm s}$ on one sublattice and $-m_{\rm s}$
on the other.
To be able to calculate the staggered susceptibility we add a
staggered-magnetic-field
term to the Hamiltonians ${\cal H}$ and ${\cal H}_{\rm SB}$:
\be
{\cal H}_{\rm mag,s} = - \sum_{i \sigma} \, h_{i,s}\sigma n_{i \sigma} ~,
\label{eq:Hmags}
\ee
where $h_{i,s}$ equals $h_{\rm s}$ on one sublattice and $-h_{\rm s}$
on the other.
The saddle-point approximation of time- and position-independent Bose-fields
on each of the two sublattices separately (introducing staggered
Lagrange multipliers as well)
results in a $2 \times 2$
problem (for each $\sigma$ separately, only one-electron states with $\bk$ and
$\bk + \bQ$ couple, $\bQ \equiv (\pi,\pi)$)
for the quasi-particle spectrum, which is easily diagonalized.
Analogously to the ferromagnet, we define the ``free energy'' per site
for the antiferromagnet as:
\be
\ffi_{\rm AF} = - \frac{1}{\beta N} \ln \mbox{Tr} e^{-\beta ({\cal H}_{\rm SB}
+
{\cal H}_{\rm mag,s})} + \mu n + h_{\rm s} m_{\rm s} ~. \label{eq:fAFdef}
\ee
It is given by:
\be
\ffi_{\rm AF} = - \frac{1}{\beta N} \sum_{\bk , \sigma}\pri  \ln \left[ 1 +
\mbox{e}^{-\beta E(\bk)} \right] + Ud^2 + \mut n + \lat_{\rm s} m_{\rm s} ~,
\label{eq:fAF}
\ee
where
\be
E(\bk) = \pm \sqrt{q_{\rm s}^2 t^2(\bk) + \lat_{\rm s}^2} - \mut ~,
\label{eq:Ek}
\ee
and we have again introduced an effective chemical potential $\mut$ and
an effective magnetic field $\lat_{\rm s}$.
The prime indicates that the sum over $\bk$ is over the magnetic Brillouin zone
only ($k_x \pm k_y \in [ -\pi ,\pi ] $).
The band-renormalization factor $q_{\rm s}$ is now a $\sigma$-independent
quantity because of the staggering:
\be
q_{\rm s} (n, m_{\rm s}, d) \equiv \langle
z^{\dagger}_{j\sigma} z_{i\sigma} \rangle_{\rm AF} =
z_{A\sigma}z_{B\sigma} = z_\uparrow z_\downarrow ~, \label{eq:qs}
\ee
where
\be
z_\sigma =
\frac{\sqrt{ (1-n+d^2)(n+\sigma m_{\rm s} - 2d^2)} +
      d \sqrt{n-\sigma m_{\rm s} - 2d^2}}
    {\sqrt{(n+\sigma m_{\rm s}) \left( 1 - \frac{n + \sigma m_{\rm s}}{2}
\right)} } ~.
\label{eq:zsig}
\ee

As for the ferromagnet, the free energy can be expressed as an integral over
the DOS of freely hopping electrons, $\Ne$:
\be
\ffi_{\rm AF} = - \frac{1}{\beta} \int \mbox{d}\eps \, \Ne \ln \left[ 1 +
\mbox{e}^{-\beta E(\eps)} \right] + Ud^2 + \mut n + \lat_{\rm s} m_{\rm s} ~.
\label{eq:fAFN}
\ee
For convenience we restrict ourselves to densities $n \leq 1$; in that case
only
the negative square root in (\ref{eq:Ek}) is relevant and we have in
(\ref{eq:fAFN}):
\be
E(\eps) =  - \sqrt{q_{\rm s}^2 \eps^2 + \lat_{\rm s}^2} - \mut ~.
\label{eq:Eeps}
\ee
The consistency equations for the antiferromagnet are obtained in the same
way as for the ferromagnet and read:
\begin{eqnarray}
U & = & \frac{q_{{\rm s}d}}{2d} \eb  ~, \label{eq:ceUAF} \\
n & = & \int \mbox{d}\eps \, \Ne f[E(\eps)] ~, \label{eq:cenAF} \\
m_{\rm s} & = & \lat_{\rm s} \int \mbox{d}\eps \,
\frac{\Ne f[E(\eps)]}{\sqrt{q_{\rm s}^2 \eps^2 + \lat_{\rm s}^2}} ~,
\label{eq:cemAF} \\
\lat_{\rm s} & = & h_{\rm s} + q_{{\rm s}m_{\rm s}} \eb ~, \label{eq:celaAF} \\
\mut & = & \mu + q_{{\rm s}n} \eb ~, \label{eq:cemuAF}
\end{eqnarray}
where we have defined:
\be
\eb = q_{\rm s} \int \mbox{d}\eps \,
\frac{\Ne \eps^2 f[E(\eps)]}{\sqrt{q_{\rm s}^2 \eps^2 + \lat_{\rm s}^2}}  ~.
\label{eq:epsav}
\ee
Expressions for partial derivatives $q_{{\rm s}\alpha}$ of $q_{\rm s}$ are
given in Appendix C.
$\ffi_{\rm AF}$ is a function of the five variables $n, m_{\rm s}, d, \mut$
and $\lat_{\rm s}$,
whereas $\mu$ and $h_{\rm s}$ are control parameters.
Calculating $\ffi_{\rm AF}$ for fixed $U/t$ and $n$, means that the
last consistency equation (\ref{eq:cemuAF})
becomes irrelevant again and the four remaining variables must be
found self-consistently from the four remaining consistency equations.
To compute the free energy the staggered field $h_{\rm s}$ is taken to be
zero. This free
energy may be compared to the free energies of the three phases discussed in
the
previous subsection and lines of first-order transitions in the $(t/U,n)$-plane
may be found (see Section 3). We note that
the equations for the paramagnet can also be found from the above equations for
the antiferromagnet by putting $m_{\rm s} = 0$ if $h_{\rm s} = 0$
(then also $\lat_{\rm s} = 0$).
In order to find the transition line for the continuous phase
transition between antiferromagnet and paramagnet an expression
for the staggered susceptibility $\chi_{\rm s}$ is derived in Appendix D.

\subsection{Spiral Phase}
To obtain the spiral phase the magnetization vector is assumed to vary
in space as:
\be
{\bf m}_i = m \left( \cos (\bq \cdot {\bf R}_i ) ,
\sin (\bq \cdot {\bf R}_i ) , 0 \right) ~. \label{eq:msp}
\ee
In Ref.\cite{FWI}, the spin-rotation invariant formulation is used to compute
the free energy  $\ffi_{\rm sp}$ for this phase. For details of this
calculation we refer to Refs.\cite{LWH,FWI}, here we suffice by quoting
the result (in our notation):
\be
\ffi_{\rm sp} = - \frac{1}{\beta N} \sum_{\bk , \nu}\, \ln \left[ 1 +
\mbox{e}^{-\beta E_{\bq,\nu}(\bk)} \right] + Ud^2 + \breve{\mu} n -
\lambda^{(2)} m~.
\label{eq:fsp}
\ee
where
\begin{eqnarray}
{}~~& \hspace{-3.15in}E_{\bq,\nu}(\bk) =  \left( z_+^2 + z_-^2 \right) \left[
\frac{t(\bk) + t(\bk + \bq)}{2} \right] - \breve{\mu} +  \nonumber \\
{}~~&  +  \nu \left\{ \left( z_+^2 - z_-^2 \right)^2
\left[ \frac{t(\bk) - t(\bk + \bq)}{2} \right]^2 +
\left[ z_+ z_-[t(\bk) + t(\bk + \bq)] + \lambda^{(2)} \right]^2
\right\}^{\frac{1}{2}}
{}~, \label{eq:Eqnuk}
\end{eqnarray}
with $\nu = \pm 1$.
The parameters $z_\pm$ are functions of $n,m$ and $d$; in terms of
the $z_\sigma$
(formula (\ref{eq:zsig}) with $m_{\rm s}$ replaced by $m$) they are given by:
\be
z_{\pm} = \ha \left( z_{\uparrow} \pm z_{\downarrow} \right) ~. \label{eq:zpm}
\ee
The parameter $\breve{\mu} = \mu - \lambda_0^{(2)}$ is again an effective
chemical potential, whereas $\lambda_0^{(2)}$ and $\lambda^{(2)}$ are the
Lagrange multipliers arising from the constraint (\ref{eq:Q2}) when made
spin-rotation invariant.\footnote{In the spin-rotation invariant formulation,
constraint (\ref{eq:Q2}) gives rise to a scalar Lagrange multiplier
$\lambda_0^{(2)}$ as well as a vector Lagrange multiplier
$\vec{\lambda}^{(2)}$.
For a spiral phase the latter results in another scalar $\lambda^{(2)}=
|\vec{\lambda}^{(2)}|$
because it must show the same spatial variation as ${\bf m}$ in
(\ref{eq:msp}).}
We did not include any explicit magnetic field in the Hamiltonian in
studying the spiral phases.
One may verify that for $\bq = (0,0)$ and $\bq = (\pi,\pi)$ (\ref{eq:Eqnuk})
reduces to the expressions (\ref{eq:Eksig}) and (\ref{eq:Ek}) for the
ferromagnet and antiferromagnet, respectively. In terms of the $z_\pm$ the
band-renormalization factors are given by $q_\sigma = (z_+ + \sigma z_-)^2$
and $q_{\rm s} = z_+^2 - z_-^2$.

For the spiral phase we do not give the consistency equations like we did
for the ferromagnet and antiferromagnet, since we are only aiming at a simple
phase diagram which does not include the spiral phases. Moreover,
the consistency
equations and free energy cannot be expressed as one-dimensional integrals
over a density of states because of the spiraling vector $\bq$ involved.
Therefore,
the consistency equations need to be solved numerically on a finite lattice.
The regions in the phase diagram where spiral phases dominate the simple
magnetic phases were calculated in Refs.\cite{FWI} and \cite{MDF}.
In this paper,
we will only use the expressions above to derive a formula within
SBMF approximation for the helicity modulus (or: spin stiffness) and
effective hopping parameter in Section 4.

\section{Magnetic Phase Diagram}
We compute, in SBMF approximation,
the complete (i.e. all first order and continuous phase
transitions are included) ground-state magnetic phase diagram for the
Hubbard model
on a square lattice allowing for the four simple magnetic phases,
paramagnet (PM), weak ferromagnet (WF), strong ferromagnet (SF) and
antiferromagnet
(AF). In their original paper, Kotliar and Ruckenstein \cite{KR} only
calculated the
lines where the PM becomes unstable towards ferromagnetic or antiferromagnetic
ordering (continuous transitions), whereas Evans \cite{Evans}
also included some first-order transitions, but not all, so that an
incomplete picture emerged.

First, the regions in the $(4t/U,n)$-plane where ferromagnetism
(either WF or SF)
and antiferromagnetism can occur are determined, by calculating the
lines where the
homogeneous and staggered susceptibilities, $\chi$ and $\chi_{\rm s}$,
of the PM diverge. In Appendices B and D expressions for $\chi$ and
$\chi_{\rm s}$
are derived. The condition that the denominator in these expressions vanishes
(generalized Stoner criterion) provides an additional equation to be
solved in conjunction with the consistency equations for the PM
(see Section 2.1). In this way,
for fixed $n$, the additional equation fixes the (critical) $U/t$ value for
which the susceptibility diverges. The resulting lines are displayed in
Figure 1(a) and agree with previously published results \cite{KR,Evans}.

Now, using the formulae in Sections 2.1 and 2.2 all first order
phase transition lines
in the $(4t/U,n)$-diagram are computed.
Since no susceptibilities are required, $h$ and $h_{\rm s}$ are
taken to be zero. In principle,
for each of the four phases for fixed values of $U/t$ and $n$ the
energy is found
by solving the consistency equations
simultaneously.\footnote{Using the analytically
known density of states (\ref{eq:DOS}) this may be conveniently done with the
program MATHEMATICA \cite{MATH}.
Some care is required in integrating through the logarithmic
singularity of $\Ne$ in $\eps = 0$.}
For the SF and PM this problem simplifies somewhat:
for the SF the set of equations (\ref{eq:ceUF})-(\ref{eq:celaF}) is
reduced by one (since $m=n$) and for the PM we have $m=0$.
For each pair of phases one then finds a line in the $(4t/U,n)$-plane where the
two energies are equal. The results of such calculations are displayed in
Fig. 1(a). In principle there are six such lines, but
the first order PM/AF transition line coincides with the continuous
PM/AF transition.
Note however that the continuous PM/F transition and the first order
PM/WF transition differ.

Taking into account all four phases, the phase diagram
of the Hubbard model on a square lattice in SBMF approximation of Figure 1(b)
emerges, in which all interrupted lines denote first-order transitions
and the full line a continuous (PM/AF) transition. We now discuss the
phase diagram in comparison with the same phase diagram as obtained in the
Hartree-Fock approximation (HFA) and in comparison with previously published
SBMF results.

The corresponding, i.e., allowing for the same
four phases, phase diagram to Fig.1(b) in the HFA is shown in Figure 1(c).
A similar diagram was given previously
in three dimensions by Penn \cite{Penn} and in two dimensions
by Hirsch \cite{Hirsch}, but in the latter the non-monotonous behaviour of
the F/AF transition line was missed and the region of WF was not determined.
Long has given the PM/F/AF phase diagram in HFA using a constant
density of states;
in that case no extremum in the F/AF transition occurs \cite{Long}.
Another surprising feature of Fig.1(c), besides the maximum in the F/AF line,
is the fact that the WF/SF transition line is found to oscillate slightly
around the line of the continuous P/F line. The difference in these
curves is very small, but we have ascertained that it is {\em not} due
to numerical inaccuracies. Thermodynamically such behaviour is allowed; it only
means that along the P/F boundary the transition is sometimes continuous and
sometimes first order. In comparing Fig.1(b) and Fig.1(c) it should be noted
that they are topologically the same, but that because of the difference in
scale on both the density and $4t/U$ axes, the SBMF has reduced the
magnetically
ordered regions considerably with respect to the HFA. Furthermore, the region
where WF dominates has grown at the expense of the SF phase. In Table I, we
list
the location of the ``tripod'' points (i.e., points where three phases meet;
the AF/SF/WF and PM/SF/WF points are triple points, i.e., points where three
first-order transitions meet), as well as the renormalization factor obtained
in
going from the HFA to the SBMF approximation. Globally speaking this factor is
about 3 for the hole density $\delta$ and also about 3 for $t/U$ (if the
ferromagnetic region is considered as a whole). Since the HFA
overestimates the importance of magnetic ordering \cite{Long}, the SBMF
approximation is clearly an improvement. The critical hole density above which
antiferromagnetism cannot occur is determined by the continuous PM/AF
transition
and is given by $\delta_{\rm c}^{\rm AF} = 0.21$ in SBMF.
More interestingly, the
critical hole density above which ferromagnetism cannot occur is in SBMF
determined by the first order PM/SF transition and given
by exactly $\delta_{\rm c}^{\rm F} = 1/3$, as a simple argument can show
(see e.g. Ref.\cite{MDF}). The latter value agrees very well with the result
$\delta_{\rm c}^{\rm F} = 0.29$ obtained from calculations using a variational
wave function \cite{vdLEd}. Remarkably, also high-temperature series expansions
for the Hubbard model find for $U/t \rightarrow \infty$ a value of about 0.33,
below which ferromagnetic nearest-neighbour correlations occur \cite{DtH},
and a value ``near 3/11'' ($\simeq 0.27$), below which a strong separation
of energy scales for spin and translational degrees of freedom is observed
\cite{Yed}.  Although these features in the high-temperature series expansions
appear to be temperature independent over a wide temperature range,
extrapolation to $T=0$ is cumbersome in such expansions \cite{DtH2}.
We also note that the $\chi^{-1} = 0$ line is nowhere in the diagram a phase
boundary; therefore the continuous PM/F transition, in this approximation and
contrary to the HFA result, is preempted by first order PM/WF or PM/SF
transitions.

A calculation of the phase diagram similar to ours was previously performed
by Evans \cite{Evans}. However, the more cumbersome PM/WF and AF/WF
first order transitions were not computed and for the WF/SF transition
only the limit $m=n$ and $d=0$ in the WF was taken. The latter determination
turns out only to give an upper bound (in $t/U$, for fixed $n$)
for the WF/SF first-order transition computed by comparing energies, as
it should. As a result the final phase diagram of Ref.\cite{Evans} is obtained
by removing from Fig.1(a) the PM/WF and AF/WF lines and replacing the
WF/SF line by one extending from (0.15,0.0) to (0.0,0.38) in the
$(4t/U,\delta)$-plane ($\delta = 1-n$).
Evans then appears to call WF only the tiny, triangle-shaped, region
at the center of
our WF region (although this is not very clear from Fig.1 in Ref.\cite{Evans}).
This assignment, however, is thermodynamically not justified
since two of the boundaries then correspond to PM/SF and PM/AF transitions.
Also the third boundary in that case (PM/F) is not a boundary for the
WF region as computed by us.

To conclude the discussion of the phase diagram in Fig.1,
we stress that the actual ground-state
phase diagram of the Hubbard model, even when constructed within either
the HF or SBMF approximation, will also have to include inhomogeneous phases
like domain walls and spiral phases. For instance, it was shown within the
SBMF approximation that
spiral phases supersede the antiferromagnet immediately when going off
half-filling
and also the ferromagnetic phases shrink somewhat in favour of
certain spiral phases \cite{FWI,MDF}. Our detailed determination of the
simple phase diagram only serves to study the consequences of the
SBMF approximation when compared with the HFA.

\section{Helicity Modulus and Effective Hopping}
A crucial quantity in the study of quantum-ordered states is the helicity
modulus, which is the stiffness associated with a twist of the order parameter,
or, equivalently, with phase fluctuations of a complex order parameter.
For the attractive Hubbard model, which exhibits superconducting or superfluid
order, the helicity modulus corresponds to the superfluid density, whereas for
the repulsive Hubbard model it is the spin stiffness
of the AF ordered phase at half-filling. In previous papers,
the helicity modulus, denoted by $\rho_{\rm s}$, was calculated for the
2D Hubbard
model both in the
HFA as well as by variational Monte Carlo methods \cite{PDrhos,PDEPL}.
Also a comparison with exact diagonalization and Quantum Monte Carlo results
was made, showing that the HFA renders quantitatively reasonable results
for $\rho_{\rm s}$ \cite{PDBR}.
For the repulsive Hubbard model this could only be shown
at half-filling. Since we have already seen that the SBMF approximation is an
improvement over the HFA, it is of interest to see what it will give for
$\rho_{\rm s}$.
In this section, we first derive an expression for $\rho_{\rm s}$
within the SBMF
approximation and use it to compute $\rho_{\rm s}$. Then the connection between
$\rho_{\rm s}$ and the effective hopping parameter is discussed, leading to
calculations of the effective hopping both at half-filling and off
half-filling.
The SBMF results are compared to results from the HFA and
Quantum Monte Carlo results.

To obtain $\rho_{\rm s}$, we can make use of the results for the spiral phase
in Section 2.3. In the AF phase the order parameter is the staggered
magnetization. This can be viewed as a spiral phase with spiral vector
$\bq = \bQ \equiv (\pi, \pi)$. A small twist in the AF order parameter
then corresponds
to a spiral vector which deviates slightly from $\bQ$:
\be
\bq = (\pi,\pi) - \tilde{\bq} ~. \label{eq:qtilde}
\ee
The helicity modulus $\rho_{\rm s}$ is given by:
\be
\rho_{\rm s} = \lim_{\tilde{q} \rightarrow 0}
\frac{\ffi(\tilde{q}) - \ffi(0)}{\ha \tilde{q}^2} ~, \label{eq:rhos1}
\ee
with the free energy per site $\ffi$ given by (\ref{eq:fsp}) and
where $\tilde{q}$ is the modulus of $\tilde{\bq}$.
To facilitate the computation it is advantageous to perform a further
manipulation.
Since the spectrum is periodic in reciprocal space and we are going to
integrate over the full Brillouin zone, it is allowed to shift the
spectrum over $\ha \tilde{\bq}$, so that (\ref{eq:Eqnuk}) becomes
(if in turn we rename $\ha \tilde{\bq}$, for notational convenience, to $\bq$):
\begin{eqnarray}
{}~~& \hspace{-2.9in}E_{\bq,\nu}(\bk) =  \left( z_+^2 + z_-^2 \right) \left[
\frac{t(\bk + \bq) - t(\bk - \bq)}{2} \right] - \breve{\mu} +  \nonumber \\
{}~~&  +  \nu \left\{ \left( z_+^2 - z_-^2 \right)^2
\left[ \frac{t(\bk + \bq) + t(\bk - \bq)}{2} \right]^2 +
\left[ z_+ z_-[t(\bk + \bq) - t(\bk - \bq)] + \lambda^{(2)}
\right]^2 \right\}^{\frac{1}{2}}~, \label{eq:Eqnuk2}
\end{eqnarray}
with $\nu = \pm 1$.
After this manipulation, the occurring sum $t(\bk + \bq) + t(\bk - \bq)$
is even in the small parameter $q$ (=$|\bq|$) and the difference is odd. The
same is not true for the small parameter $\tilde{q}$ in the occurring
sum and difference $t(\bk) \pm t(\bk + \bQ - \tilde{\bq})$ in (\ref{eq:Eqnuk}).
In this notation, $\rho_{\rm s}$ is given by:
\be
\rho_{\rm s} = \lim_{q \rightarrow 0} \frac{\ffi(q) - \ffi(0)}{2q^2} ~,
\label{eq:rhos2}
\ee
If we now restrict to ground-state properties ($T=0$) and
densities less than half-filling ($n < 1$), we only need to expand
the $\nu = -1$-branch for small $q$ (and integrate over the Brillouin zone
(BZ))
to obtain $\ffi(q) - \ffi(0)$ to order $q^2$. We find (taking $\bq = (q,0)$ for
convenience) that the term proportional to $q$ vanishes after integrating over
BZ, and that $\rho_{\rm s}$ is given by:
\be
\rho_{\rm s} = - \frac{t q_{\rm s}^2}{N_{\rm s}} \sum_{\bk} \, \left[
\frac{t(\bk) \cos (k_x)}{2E(\bk)} +
\frac{4t z_+^2z_-^2 t^2(\bk) \sin^2 (k_x)}{E^3(\bk)} \right] \label{eq:rhosk}
\ee
with
\be
E(\bk) = \sqrt{q_{\rm s}^2 t^2(\bk) + \lambda^2} ~, \label{eq:Ek2}
\ee
and the band renormalization $q_{\rm s}$ is:
\be
q_{\rm s} = z_+^2 - z_-^2 ~. \label{eq:qszpzm}
\ee
We have omitted the superscript $(2)$ on $\lambda$. We remark that if the above
shift in the BZ is not performed, a much longer expression for $\rho_{\rm s}$
results; the expression is equivalent to (\ref{eq:rhosk}), but this is not
trivial. We further note that the $T=0$ HFA result of Ref.\cite{PDrhos} is
recovered by omitting the second term in the BZ sum and putting $q_{\rm s}$
equal to 1.

In order to compute $\rho_{\rm s}$ for fixed density $n$,
according to Section 2.2, the following set
of equations needs to be solved self-consistently
(for $T=0$ and $h_{\rm s} = 0$;
cf. (\ref{eq:ceUAF})-(\ref{eq:celaAF})):
\begin{eqnarray}
m_{\rm s} & = & 2 \lab \int_{\mub}^{4} \mbox{d}\eps \,
\frac{\Ne}{\left(\eps^2 + \lab^2 \right)^{1/2}}         ~, \label{eq:cemAF2} \\
\lab & = & \frac{2}{q_{\rm s}} q_{{\rm s}m_{\rm s}} \int_{\mub}^{4}
\mbox{d}\eps \,
\frac{\Ne \eps^2}{\left(\eps^2 + \lab^2 \right)^{1/2}}     ~,
\label{eq:celaAF2} \\
U & = & \frac{q_{{\rm s}d}}{d} \int_{\mub}^{4} \mbox{d}\eps \,
\frac{\Ne \eps^2}{\left(\eps^2 + \lab^2 \right)^{1/2}}       ~,
\label{eq:ceUAF2}
\end{eqnarray}
where $\lab = \lambda/q_{\rm s}$ and
$\mub$ is determined by the fixed $n$:
\be
n = 2 \int_{\mub}^{4} \mbox{d}\eps \, \Ne ~, \label{eq:cenAF2}
\ee
In terms of the parameters of Section 2.2, we have $\mub =
\sqrt{\mut^2 - \lat^2_{\rm s}}/q_{\rm s}$
($\lat_{\rm s}$ is called $\lambda$ here).
The band renormalization $q_{\rm s}$ is given by (\ref{eq:qs})-(\ref{eq:zsig}).
In terms of an integral over the DOS, the energy of the AF state (per site)
and the spin stiffness $\rho_{\rm s}$ are given by:
\begin{eqnarray}
e_{\rm AF} = -2 q_{\rm s} \int_{\mub}^{4} \mbox{d}\eps \, \Ne \sqrt{\eps^2 +
\lab^2} + Ud^2 + \lab q_{\rm s} m_{\rm s}  \label{eq:eaf} \\
\rho_{\rm s} = \frac{q_{\rm s}}{4} \int_{\mub}^{4} \mbox{d}\eps \,
\frac{\Ne \eps^2}{\left(\eps^2 + \lab^2 \right)^{1/2}} -
\frac{z_+^2z_-^2}{q_{\rm s}}
\int_{\mub}^{4} \mbox{d}\eps \, \frac{\Nve \eps^2}{\left(\eps^2 + \lab^2
\right)^{3/2}} \label{eq:rhose}
\end{eqnarray}
where $\Nve$ is the weighted density of states:
\be
\Nve \equiv \frac{1}{N} \sum_{\bk} \, \left[ {\bf \nabla} t(\bk) \right]^2
\delta( \eps - t(\bk) ) ~, \label{eq:Nvedef}
\ee
which for a square lattice can be calculated analytically
(see Ref.\cite{BelkR}):
\be
\Nve = \frac{8t}{\pi^2} \left\{ E\left[ 1 - \left(\frac{\eps}{4t}\right)^2
\right]
-\left(\frac{\eps}{4t}\right)^2 K\left[ 1 - \left(\frac{\eps}{4t} \right)^2
\right] \right\} ~, \label{eq:NveEK}
\ee
for $|\eps| \leq 4t$ and zero otherwise. $K(x)$ and $E(x)$ are the
complete elliptic
integrals of the first and second kind, respectively. We remark that using the
weighted density of states the finite-temperature result for $\rho_{\rm s}$
obtained in the HFA \cite{PDrhos} can also be written as an integral over
energy.

Since only at half-filling the AF phase is the ground-state, we first restrict
ourselves to this case: $n=1$ ($\mub = 0$).
In Table II, self-consistent parameters
are given for various $U/t$; the corresponding energy $e_{\rm AF}$
and band renormalization $q_{\rm s}$ are also given. These results agree with
those published previously by Hasegawa \cite{Has}.
In particular, we note that $q_{\rm s}$ never deviates from 1 by more than 5
\%,
implying that slave bosons at half-filling renormalize the Hartree-Fock results
only by a small amount.
As concerns $\rho_{\rm s}$, for the case of half-filling the integral
containing
$\Nve$ plays no role since for $n=1$ we have $z_- = 0$ (as is easily verified
from (\ref{eq:zsig}) and (\ref{eq:zpm})).
In Table III and Figure 2, we compare the results for $\rho_{\rm s}$ obtained
in the SBMF, with those obtained previously using the HFA and using variational
Monte Carlo calculations with an (antiferromagnetic) Gutzwiller wave function
(GWVMC) \cite{PDrhos}. The fact that slave bosons only renormalize
the HF results a
little bit is reflected in the fact that the SBMF and HF results never differ
by more than 7 \%. Note however that the SBMF result is always larger
(except for very small $U/t$, $U/t < 2.5$), whereas
$q_{\rm s}$, which enters as a factor in (\ref{eq:rhose}), is smaller than 1.
The direct
effect of $q_{\rm s}$ is more than compensated by a renormalized
(smaller) value of the antiferromagnetic gap $\lambda = \lab q_{\rm s}$.
A further observation from Table III and Fig.2 is that the SBMF results almost
coincide with the GWVMC results. The difference between the results is an
indication of the difference between the Gutzwiller {\em approximation} (which
is equivalent to the present SBMF approximation, see the introduction) and
the Gutzwiller {\em wave function}. Although these are not identical
\cite{Voll},
the difference for the spin stiffness is not big, as shown in Fig.2.
Therefore the tedious variational Monte Carlo calculations for
$\rho_{\rm s}$ can
be replaced by the above set of equations which are exact (within the SBMF)
and easy to solve.

On very general grounds it can be derived that the helicity modulus
(spin stiffness
for positive $U$ and superfluid weight for negative $U$) comprises of
a ``direct''
part proportional to the average kinetic energy $\langle T \rangle$
and a part related to the current-current correlation function $\Lambda_{xx}$
\cite{ScalWZ}.
The HFA effectively neglects the $\Lambda_{xx}$-part, whereas GWVMC
calculations
find only a negligible correction to the kinetic part
$-{\scriptstyle \frac{1}{8}}\langle T \rangle$
\cite{PDrhos}. Since in formula (\ref{eq:rhosk}) the first term is exactly
$-{\scriptstyle \frac{1}{8}}\langle T \rangle$ in the SBMF approximation and
at half-filling the second term equals zero, we can conclude that
also the SBMF approximation only gives the kinetic part of $\rho_{\rm s}$.
In Ref.\cite{PDBR} it was estimated (by comparing to appropriate exact
diagonalization
calculations and Quantum Monte Carlo (QMC) calculations) that the HFA
overestimates
$\rho_{\rm s}$ at half-filling by 68, 40, 38 and 36 \% for $U/t$ = 4, 8,
10 and 20,
respectively. However, if one compares the kinetic energy found in the HF and
SBMF approximations with QMC calculations for $n=1$ (The latter are obtained
from Ref.\cite{QMCn1}), one concludes that
the approximations perform very satisfactorily.
This is illustrated in Figure 3,
where we plot the effective hopping integral $t_{\rm eff}$, defined by
normalizing
the average kinetic energy for interaction constant $U$ with that for $U=0$:
\be
\frac{t_{\rm eff}}{t} = \frac{\langle c_{i \sigma}^{\dagger} c_{j \sigma} +
                              c_{j \sigma}^{\dagger} c_{i \sigma} \rangle_U}
                {\langle c_{i \sigma}^{\dagger} c_{j \sigma} +
                         c_{j \sigma}^{\dagger} c_{i \sigma} \rangle_{U=0}}  ~.
\label{eq:teff}
\ee
The denominator is easily evaluated as the energy of the PM phase
in the HFA, since for $U=0$
the HFA is exact and there is no potential energy.
The QMC data is taken at sufficiently low temperature ($\beta t = 16$) for this
comparison with $T=0$ results to be meaningful.
We note that a similar comparison of SBMF and QMC data was made in
Ref.\cite{Lil};
in that paper the SBMF approach was formulated as a 14-dimensional
optimization problem. By comparing their Figure 2(b) with our Fig.3, the
results
from our simple formula (\ref{eq:rhose}) are found to be the same.
A surprising feature of Fig.3 is perhaps not so much that the SBMF results
approximate the QMC results so well, but that the HFA results do the same
already.

We now discuss the off-half-filling case; since off half-filling their exists a
spiraling vector ${\bf q}_0$ for which the spiral phase has lower energy
than the
AF phase, we cannot call $\rho_{\rm s}$ as given by (\ref{eq:rhose})
the stiffness
of the ground state anymore. Instead, the expression (\ref{eq:rhose}) has the
interpretation of the stiffness of the AF phase with respect to a small
deviation
from spiraling vector $(\pi,\pi)$. In order to compute the stiffness of
the ground
state one would have to perturb the spiral phase with vector ${\bf q}_0$, but
this is beyond the scope of the present paper. Here we only investigate how
well the effective hopping (or, equivalently, the kinetic energy) of the
two-dimensional Hubbard model off half-filling is described by the first
term in (\ref{eq:rhose}). In Figure 4(a), we compare $t_{\rm eff}/t$
obtained from Hartree-Fock and SBMF approximations (for $T=0$) with
low-temperature ($\beta t = 6$) QMC data
(The latter are obtained from Ref.\cite{QMCnneq1}). The results are displayed
as a function of density $n$ for the one value of $U/t$ (=4) for which there
are
QMC data available. From the phase diagrams in Fig.1 it is clear that both
in the HFA and SBMF approximation an AF/PM transition
occurs\footnote{If the restricted set of four phases is considered as before.}
for some critical
density $n_{\rm c}$ ($n_{\rm c}=0.86$ in SBMF and $n_{\rm c}=0.76$ in HFA
for $U/t=4$, the former is not showing in Fig.1(a)).
Below $n_{\rm c}$ (paramagnetic phase), $t_{\rm eff}/t$ equals 1 in the HFA
and equals the band renormalization $q$ (=$q_{\uparrow} = q_{\downarrow}$)
in the SBMF approximation.
Clearly the SBMF approach is a significant improvement over the HFA;
the agreement
with the QMC data is less good in the density interval just off half-filling.
This is most probably caused by the fact that for these densities the
assumed (antiferromagnetic) phase in the SBMF approach is not the correct one.
The same remark concerning Ref.\cite{Lil} as made above for $n=1$ is
appropriate here.
Finally, in Figure 4(b), we also show the results for the
effective hopping integral
for a few other values of $U/t$, for which no QMC data are available.
The corresponding HFA curves are not shown, but the behaviour is similar as for
$U/t = 4$: at half-filling $t_{\rm eff}/t$ is somewhat below the SBMF result
and rises to 1 in going off half-filling.
The densities below which
$t_{\rm eff}/t$ equals 1 can be read off from Fig.1(c). We note that in
Fig.4(b)
the non-differentiability at $n_{\rm c}$ (which is close to 0.8 for
$U/t = 8, 12, 16$,
as can be seen in Fig.1(a) and (b)) is less pronounced for
$U/t = 8, 12, 16$ than it is for $U/t=4$.

\section{Discussion and Conclusions}
Above we have given a detailed account of calculations within the slave-boson
mean-field (SBMF) approximation for the repulsive Hubbard model on a
square lattice.
We have focused on the phase diagram, a particular response function,
the helicity
modulus $\rho_{\rm s}$, and the related effective hopping integral.
All calculations can be expressed in terms of a set of integral equations
with one-dimensional
integrals over energy containing a density of states, which is known
analytically
for the square lattice. These equations are solved self-consistently.

If for the ($t/U,n$)
phase diagram we restrict to simple magnetic phases, the SBMF approach
is found to reduce the magnetically ordered regions with respect
to the Hartree-Fock
approximation (HFA). Along the density axis the reduction is roughly
a factor of 3,
whereas along the $t/U$-axis the reduction of the ferromagnetic region
(weak- and strong ferromagnetism together) is also a factor of 3. In the SBMF
approach the portion of weak ferromagnetism grows at the expense of the
strong-ferromagnetism portion when compared to the HFA. The present SBMF phase
diagram is more likely to be a good starting point for more
sophisticated approaches to the phase diagram than is the HFA phase diagram,
because in an infinite number of dimensions the SBMF approach becomes exact
(see the introduction).
On the other hand, it is seen that the SBMF approach is not qualitatively
different from the HFA, but rather a renormalized form of HFA.

The new quantity that we obtain within the SBMF approach is the helicity
modulus
$\rho_{\rm s}$. At half-filling, the results for $\rho_{\rm s}$ practically
coincide with those obtained using variational Monte Carlo calculations
with a Gutzwiller wave function and are
generally somewhat larger (about 5 \%) than those
obtained in the HFA. The exact results are estimated to be smaller than the HFA
results. This discrepancy is due to the neglect within the SBMF approach
(as in the HFA) of the current-current correlation part of $\rho_{\rm s}$.
The remaining {\em kinetic part} of $\rho_{\rm s}$ agrees very well with
Quantum Monte Carlo (QMC) calculations of the kinetic energy or
effective hopping
integral. Also off half-filling, where our expression for $\rho_{\rm s}$
no longer has the interpretation of helicity modulus of the ground state,
it is found to represent the effective hopping much better than the HFA.

A number of extensions of the present work is possible.
Most of the extensions discussed below have already been performed
within the HFA \cite{PDrhos} and since the SBMF approach turns out to
be a renormalized form of the HFA, the qualitative effect of such
expansions on the present SBMF results can be predicted.
A possible sequel to the present work (which was not attempted before
for the HFA) is the calculation of the
helicity modulus for the spiral or domain-wall phases, which supersede
the AF ground state that one has at half-filling and which is the
starting point for our calculated helicity modulus.
Another extension is to introduce a homogeneous magnetic field $h$ in the
AF phase at half-filling. Using the well-known
mapping between repulsive and attractive
Hubbard models (see e.g. Ref.\cite{PDrhos}) the corresponding expression for
$\rho_{\rm s}$ than equals the superfluid density of the attractive ($U<0$)
Hubbard model off half-filling (without a magnetic field).
A final straightforward but tedious extension of the present results is to
allow for finite temperatures. The formalism set up above is
perfectly capable of dealing with this more general case, but the
calculations become somewhat more tedious than for $T=0$.

\section*{Acknowledgments}
We acknowledge discussions with J.M.J. van Leeuwen, J. Zaanen,
and M.L. Horbach
on various aspects of the work presented in this paper.

\newpage
\appendix
\section*{Appendix A: Band-renormalization factor $q_\sigma$}
\renewcommand{\theequation}{A.\arabic{equation}}
\setcounter{equation}{0}
In this appendix expressions for partial derivatives of the
band-renormalization
factor $q_\sigma$ for the ferromagnet are given. The formula for $q_\sigma$,
which
is a function of density $n$, magnetization $m$, and density of doubly occupied
sites $d^2$, is repeated here (see (\ref{eq:qsig})):
\be
q_\sigma (n,m,d) \equiv \langle z^{\dagger}_{j\sigma}z_{i\sigma} \rangle =
\frac{\left[ \sqrt{ (1-n+d^2)(n+\sigma m -2d^2)} + d \sqrt{n-\sigma m -2d^2}
\right]^2}{ (n+\sigma m) \left[ 1 - \ha(n + \sigma m) \right] } ~.
\label{eq:qsigA}
\ee
For $n, m$ and $d$ in the physically relevant range (e.g., $m$ should be less
or equal than $n$ and $d^2$ should be less or equal than $\ha n$) $q_\sigma$
attains values between 0 and 1, so the free electron bands are
{\em narrowed} in the SBMF approximation.

First partial derivatives $q_{\sigma n}$, $q_{\sigma m}$ and $q_{\sigma d}$
with respect to $n, m$ and $d$, respectively, are:
\begin{eqnarray}
q_{\sigma n} & = & \frac{N_\sigma}{D_\sigma} \left[ \frac{e}{r_\sigma} -
\frac{r_\sigma}{e} + \frac{d}{r_{-\sigma}} \right] -
\frac{N_\sigma^2}{D_\sigma^2} (1 - n - \sigma m) \\
q_{\sigma m} & = & \sigma \frac{N_\sigma}{D_\sigma} \left[ \frac{e}{r_\sigma} -
\frac{d}{r_{-\sigma}} \right] - \sigma \frac{N_\sigma^2}{D_\sigma^2}
(1 - n - \sigma m) \\
q_{\sigma d} & = & 2d \frac{N_\sigma}{D_\sigma} \left[ \frac{r_\sigma}{e} -
\frac{2e}{r_\sigma} + \frac{r_{-\sigma}}{d} - \frac{2d}{r_{-\sigma}} \right] ~,
\end{eqnarray}
where we have introduced the abbreviations:
\begin{eqnarray}
N_\sigma & = & \sqrt{(1-n+d^2)(n+\sigma m - 2d^2)} +
d \sqrt{n- \sigma m - 2d^2} \\
D_\sigma & = & n + \sigma m - \ha ( n + \sigma m )^2 \\
e & = & \sqrt{1 - n + d^2} \\
r_\sigma & = & \sqrt{n + \sigma m -2d^2}
\end{eqnarray}

To determine where the susceptibility diverges (see Appendix B),
we also require $q_{mm}$,
the second derivative of $q_\sigma$ with respect to $m$ calculated at $m=0$
(which is independent of $\sigma$):
\be
q_{mm} = \frac{N_0^2}{D_0^2} \left[ 1 + \frac{2(1-n)^2}{D_0} \right] -
\frac{2(1-n)^2}{D_0^2} - \frac{2ed}{(n-2d^2)D_0} ~,
\ee
with
\begin{eqnarray}
N_0 & = & \sqrt{n-2d^2} \left( \sqrt{1-n+d^2} + d \right) \\
D_0 & = & n (1 - n/2) ~.
\end{eqnarray}

\section*{Appendix B: Homogeneous magnetic susceptibility for paramagnet}
\renewcommand{\theequation}{B.\arabic{equation}}
\setcounter{equation}{0}
In this appendix we derive a formula for the homogeneous magnetic
susceptibility $\chi$ in the paramagnetic phase:
\be
\chi \equiv \left( \frac{\partial m}{\partial h} \right)_{h=0}  ~.
\ee
The derivation proceeds as follows: in the paramagnet, we have $m=0$ if $h=0$
in
the equations (\ref{eq:ceUF})--(\ref{eq:cemuF})
of Section 2.1. Then also $\lab = 0$. The solutions of the consistency
equations of the remaining variables we call $n_0$, $d_0$ and $\mub_0$. Now we
apply an infinitesimal magnetic field $\delta h$. Then $m$ and $\lab$ will
acquire small non-zero values $\delta m$ and $\delta \lambda$ and the other
quantities will deviate slightly from their values for $h=0$, because all
are coupled through (\ref{eq:ceUF})--(\ref{eq:cemuF}).
If we now work at a fixed density
(i.e., $n$ is not allowed to deviate from $n_0$ and equation (\ref{eq:cemuF})
becomes irrelevant), we have four equations containing
five small quantities. From these the required ratio
$\frac{\partial m}{\partial h}$
is obtained. Working out this procedure, by expanding all equations to first
order in the small quantities, it turns out (perhaps not surprisingly)
that the equations for $\delta \mub$ and $\delta d$ decouple from those
for $\delta m$, $\delta \lambda$ and $\delta h$. Here we give the
equations for the latter three quantities for the case $T=0$ (for which the
derivative of the Fermi-Dirac distribution is a convenient delta-function):
\begin{eqnarray}
\delta m & = & a \delta m + \chi_0 \delta \lambda  ~, \label{eq:delm} \\
\delta \lambda & = & \delta h + b \delta m + a \delta \lambda ~,
\label{eq:dell}
\end{eqnarray}
where we have introduced the following notation:
\begin{eqnarray}
\chi_0 & = & \frac{2 {\cal N}_{\rm F}}{q} \\
a & = & - \frac{2 {\cal N}_{\rm F} q_m \mub_0}{q^2} \\
b & = & -2q_{mm}\eb_0 + \frac{2 {\cal N}_{\rm F} q_m^2 \mub_0^2}{q^3} ~,
\label{eq:bchi}
\end{eqnarray}
with
\begin{eqnarray}
{\cal N}_{\rm F} & = & {\cal N}(\mub_0/q) \\
\eb_0 & = &  \int_{-\infty}^{ \mub_0/q} \mbox{d}\eps \, \Ne \eps
\end{eqnarray}
and $q$ and $q_m$ are the functions $q_\sigma$ and $q_{\sigma m}$ taken at
$m=0$
(see Appendix A).
Solving (\ref{eq:delm}) and (\ref{eq:dell}) for $\chi$ finally gives
the required formula for the susceptibility:
\be
\chi = \frac{\chi_0}{(1-a)^2 - b \chi_0} ~. \label{eq:chif}
\ee
A similar result was given in Ref.\cite{Evans}, although there is a factor of 2
difference in the $\eb_0$-term in (\ref{eq:bchi}).

\section*{Appendix C: Band-renormalization factor $q_{\rm s}$}
\renewcommand{\theequation}{C.\arabic{equation}}
\setcounter{equation}{0}
In this appendix expressions for partial derivatives of the
band-renormalization
factor $q_{\rm s}$ for the antiferromagnet are given. The formula for
$q_{\rm s}$, which
is a function of density $n$, staggered magnetization parameter $m_{\rm s}$,
and
density of doubly occupied sites $d^2$, is repeated here (cf.
(\ref{eq:qs})-(\ref{eq:zsig})):
\be
q_{\rm s} (n,m_{\rm s},d) = z(n,m_{\rm s},d) z(n,-m_{\rm s},d) ~,
\label{eq:qsC}
\ee
where
\be
z(n,m_{\rm s},d) =
\frac{\sqrt{ (1-n+d^2)(n + m_{\rm s} - 2d^2)} +  d \sqrt{n - m_{\rm s} - 2d^2}}
 {\sqrt{(n + m_{\rm s}) \left( 1 - \frac{n + m_{\rm s}}{2} \right)} } ~.
\label{eq:zsigC}
\ee
Introducing the abbreviations $N_\pm$ and the partial derivative of $z$
with respect to $m_{\rm s}$:
\begin{eqnarray}
N_\pm & = &\sqrt{(n \pm m_{\rm s} ) \left( 1 - \frac{n \pm m_{\rm s}}{2}
\right) },\\
\frac{\partial z}{\partial m_{\rm s}} & = & \frac{1}{2N_+} \left\{
\sqrt{\frac{1-n+d^2}{n+m_{\rm s}-2d^2}} - \frac{d}{\sqrt{n-m_{\rm s}-2d^2}} -
\frac{z(n,m_{\rm s},d)}{N_+} \left[ 1 - n - m_{\rm s} \right] \right\} ~~~~~~~
\end{eqnarray}
the first partial derivatives with respect to $m_{\rm s}$ and $d$ are (since we
always work at fixed density, the derivative with respect to $n$ is not
needed):
\begin{eqnarray}
q_{{\rm s}m_{\rm s}} & = &
z(n,-m_{\rm s},d) \frac{\partial z(n,m_{\rm s},d)}{\partial m_{\rm s}} +
z(n,m_{\rm s},d) \frac{\partial z(n,-m_{\rm s},d)}{\partial m_{\rm s}} \\
q_{{\rm s}d} & = & \frac{4d}{N_+ N_-} \left\{
\frac{(n-2d^2)(2n-1-4d^2) - m_{\rm s}^2}{\sqrt{n^2 - m_{\rm s}^2 - 4nd^2 +
4d^4}} +
\frac{(1-n)n - 8d^4 +(8n-6)d^2}{2d \sqrt{1-n+d^2} } \right\} ~~~~~~~
\end{eqnarray}

For the calculation of the staggered susceptibility (see Appendix D) the
second derivative of $q_{\rm s}$ with respect to $m_{\rm s}$ at $m_{\rm s} = 0$
is required; the
formula is:
\be
q_{m_{\rm s} m_{\rm s}} \equiv \left( \frac{\partial^2 q_{\rm s}}{\partial
m_{\rm s}^2} \right)_{m_{\rm s} = 0} =
\frac{4n^2 - 8n + 8}{n^3 (2-n)^3} \left[ \sqrt{1-n+d^2} + d \right]^2 (n-2d^2)
-
\frac{2(1-n+2d^2)}{n(2-n)(n-2d^2)} ~.
\ee

\section*{Appendix D: Staggered magnetic susceptibility for the paramagnet}
\renewcommand{\theequation}{D.\arabic{equation}}
\setcounter{equation}{0}
In this appendix we derive a formula for the staggered magnetic
susceptibility in the
paramagnetic phase $\chi_{\rm s}$:
\be
\chi_{\rm s} \equiv \left( \frac{\partial m_{\rm s}}{\partial h_{\rm s}}
\right)_{h_{\rm s}=0}  ~.
\ee
The procedure is as follows: the consistency equations
(\ref{eq:ceUAF})--(\ref{eq:cemuAF}) for the
antiferromagnet in Section 2.2 allow for a paramagnetic
solution in which $h_{\rm s} = 0$ leads to $m_{\rm s} = 0$ as well as
$\lat_{\rm s} = 0$.
Starting from this solution, we apply (at a fixed density)
an infinitesimal staggered magnetic field $\delta h_{\rm s}$.
This introduces small changes in the other parameters, in particular
$m_{\rm s}$ and $\lat_{\rm s}$ acquire small values
$\delta m_{\rm s}$ and $\delta \lat_{\rm s}$
(the changes in $d$ and $\mut$ are irrelevant for the present discussion).
Restricting ourselves to the $T=0$ case, the two equations relating
$\delta h_{\rm s}$, $\delta m_{\rm s}$ and $\delta \lat_{\rm s}$ are
(expanding the
consistency equations to first order in $\delta h_{\rm s}$,
$\delta m_{\rm s}$ and $\delta \lat_{\rm s}$):
\begin{eqnarray}
\delta m_{\rm s} & = & \chi_{\rm s,0} \delta \lat_{\rm s}  ~,
\label{eq:delms} \\
\delta \lat_{\rm s} & = & \delta h_{\rm s} + b_{\rm s} \delta m_{\rm s} ~,
\label{eq:dellas}
\end{eqnarray}
where the following abbreviations are introduced:
\begin{eqnarray}
\chi_{\rm s,0} & = & \frac{2}{q} \int_{-\infty}^{-\mut/q} \mbox{d}\eps \,
\frac{\Ne}{\eps} ,\\
b_{\rm s} & = & 2q_{m_{\rm s} m_{\rm s}} \int_{-\infty}^{-\mut/q}
\mbox{d}\eps \, \Ne \eps ~.
\end{eqnarray}
Here the parameter $q$ is $q_{\rm s}$ taken at $m_{\rm s} = 0$ (see Appendix
C).
Solving (\ref{eq:delms}) and (\ref{eq:dellas}) for $\chi_{\rm s}$ one has:
\be
\chi_{\rm s} = \frac{\delta m_{\rm s}}{\delta h_{\rm s}} = \frac{\chi_{\rm
s,0}}{1 - b_{\rm s} \chi_{\rm s,0}} ~.
\ee
A similar result was obtained in Ref.\cite{Evans}.

\newpage
\begin{tabular}{cccc} \hline \hline
\multicolumn{4}{c}{} \\
 $4t/U$ & SBMF & HF & HF/SBMF \\
\hline
 AF/WF/SF~~ &  0.107  &  0.551  &  5.2 \\
 AF/WF/PM~~ &  0.183  &  0.548  &  3.0 \\
 WF/SF/PM~~ &  0.057  &  0.526  &  9.2 \\
\hline
\multicolumn{4}{c}{} \\
 $\delta$ &         &         &       \\
\hline
 AF/WF/SF~~ &  0.067  &  0.25  &  3.7 \\
 AF/WF/PM~~ &  0.187  &  0.42  &  2.2 \\
 WF/SF/PM~~ &  0.275  &  0.45  &  1.6 \\
\hline
\multicolumn{4}{c}{} \\
 $\delta_{\rm c}$ &         &         &       \\
\hline
{} ~~~~AF~~~ &  0.21  &  0.42  &  2.0 \\
{} ~~~~F~~~~ &  1/3  &   1     &  3.0 \\
\hline \hline \\
\end{tabular}\\[0.5cm]
Table I. Comparison of location of tripod points (where three phases meet)
in ($4t/U, \delta$) phase diagram between Hartree-Fock approximation (HF)
and slave-boson mean-field approximation (SBMF). Also the critical hole
densities $\delta_{\rm c}$ for antiferromagnetism (AF) and ferromagnetism (F)
to occur
in both approximations are compared. The last column gives the ratio of the
HF and SBMF result in each case.\\[1.0cm]

\newpage
\begin{tabular}{cccclc} \hline \hline
\multicolumn{6}{c}{} \\
 $U/t$ & $\lab$ & $m_{\rm s}$ & $d$ & $- e_{\rm AF}$ &
$q_{\rm s}(1,m_{\rm s},d)$ \\
\multicolumn{6}{c}{} \\
\hline
\multicolumn{6}{c}{} \\
 0  &   ~~ 0       &   ~~0        &  ~~0.5       &  ~~1.62114   &   ~~ 1   \\
 1  &   ~~0.042    &  ~~0.093693  &  ~~0.478519  &  ~~1.38112   &  ~~0.994185
\\
 2  &   ~~0.24382  &  ~~0.293750  &  ~~0.444454  &  ~~1.16716   &  ~~0.980426
\\
 3  &   ~~0.55345  &  ~~0.461272  &  ~~0.404445  &  ~~0.986744  &  ~~0.966375
\\
 4  &   ~~0.94059  &  ~~0.592152  &  ~~0.364523  &  ~~0.838877  &  ~~0.956338
\\
 6  &   ~~1.89519  &  ~~0.768048  &  ~~0.292458  &  ~~0.623639  &  ~~0.951682
\\ 8  &   ~~2.98950  &  ~~0.863161  &  ~~0.235612  &  ~~0.485104  &  ~~0.959478
\\
10  &   ~~4.11613  &  ~~0.913012  &  ~~0.193926  &  ~~0.393528  &  ~~0.968953
\\
12  &   ~~5.23030  &  ~~0.940510  &  ~~0.163654  &  ~~0.330002  &  ~~0.976425
\\
16  &   ~~7.39950  &  ~~0.967337  &  ~~0.123965  &  ~~0.248782  &  ~~0.985641
\\
20  &   ~~9.51220  &  ~~0.979394  &  ~~0.099524  &  ~~0.199420  &  ~~0.990505
\\
200.1 &  ~~ 100    &  ~~0.999800  &  ~~0.009998  &  ~~0.019976  &  ~~0.999900
\\
\multicolumn{6}{c}{} \\
\hline \hline \\
\end{tabular}\\[0.5cm]
Table II. Self-consistent parameters for antiferromagnetic ground state at
half-filling in slave-boson mean-field approximation
as a function of $U/t$. The ground-state energy
for $U/t =0$ equals $-16/\pi^2$ exactly.

\newpage
\begin{tabular}{cccc} \hline \hline
\multicolumn{4}{c}{} \\
 $U/t$ & $\rho_{\rm s}^{\rm HFA}$ & $\rho_{\rm s}^{\rm SBMF}$ &
$\rho_{\rm s}^{\rm GWVMC}$  \\
\multicolumn{4}{c}{} \\
\hline
\multicolumn{4}{c}{} \\
  0  &   ~~0.2026    &  ~~0.2026   &  ~~---     \\
  1  &   ~~0.2023    &  ~~0.2012   &  ~~---     \\
  2  &   ~~0.1960    &  ~~0.1953   &  ~~0.197   \\
  3  &   ~~0.1820    &  ~~0.1847   &  ~~0.185   \\
  4  &   ~~0.1650    &  ~~0.1713   &  ~~0.172   \\
  6  &   ~~0.1332    &  ~~0.1421   &  ~~0.141   \\
  8  &   ~~0.1090    &  ~~0.1162   &  ~~0.117   \\
 10  &   ~~0.0912    &  ~~0.0962   &  ~~0.098   \\
 12  &   ~~0.0781    &  ~~0.0814   &  ~~0.083   \\
 16  &   ~~0.0602    &  ~~0.0618   &  ~~---     \\
 20  &   ~~0.0488    &  ~~0.0497   &  ~~---     \\
\multicolumn{4}{c}{} \\
\hline \hline \\
\end{tabular}\\[0.5cm]
Table III. Spin-stiffness $\rho_{\rm s}$ of the antiferromagnetic ground-state
at half-filling as a function of $U/t$ as calculated in the
Hartree-Fock approximation (HFA), the slave-boson mean-field approximation
(SBMF),
and from variational Monte Carlo calculations using the Gutzwiller wavefunction
(GWVMC). For $U/t = 0$, $\rho_{\rm s}$ equals $2/\pi^2$ exactly.
For HFA and SBMF,
results are for an infinitely large lattice, for GWVMC, results are for an
$8 \times 8$ lattice, except for $U/t = 2, 3$ which are for $20 \times 20$
and $14 \times 14$ lattices, respectively (see also Ref.\cite{PDrhos}).
\newpage
\section*{Figure captions}
\setlength{\parindent}{0.0cm}

{\bf Figure 1} Ground-state ($4t/U,\delta$) phase diagram of the Hubbard
model on a square lattice, restricted
to simple magnetic phases: paramagnet (PM), antiferromagnet (AF), weak (WF) and
strong (SF) ferromagnet. $\delta$ is the density of holes: $1-n$.
(a) Construction diagram showing all continuous and
first-order transition lines obtained in slave-boson mean-field approximation
(SBMF),
(b) phase diagram in SBMF, and (c) corresponding phase diagram in the
Hartree-Fock approximation.
Note the difference in scales of (b) and (c).\\
\vspace{1.0cm} \\
{\bf Figure 2}
Helicity modulus $\rho_{\rm s}$ for the repulsive Hubbard model on a square
lattice at half-filling as a function of $U/t$. Shown are results from the
Hartree-Fock approximation (HFA), from the slave-boson mean-field approximation
(SBMF), and from variational Monte Carlo calculations using a Gutzwiller
projected wave function (GWVMC, from Ref.\cite{PDrhos}). \\
\vspace{1.0cm} \\
{\bf Figure 3} Effective hopping integral $t_{\rm eff}/t$ for the
repulsive Hubbard model on a square
lattice at half-filling as a function of $U/t$. Shown are results from the
Hartree-Fock approximation (HFA), the slave-boson mean-field approximation
(SBMF), and Quantum Monte Carlo calculations (QMC, from Ref.\cite{QMCn1}). \\
\vspace{1.0cm} \\
{\bf Figure 4} Effective hopping integral $t_{\rm eff}/t$ for the
repulsive Hubbard model on a square
lattice as a function of electron density $n$. (a) For $U/t = 4$, results
are shown from the
Hartree-Fock approximation (HFA), the slave-boson mean-field approximation
(SBMF), and Quantum Monte Carlo calculations (QMC, from Ref.\cite{QMCnneq1}),
and (b) for $U/t = 4, 8, 12$, and 16
from SBMF calculations (QMC results for $U/t=4$ only).
The dashed lines indicate the electron densities
$n_{\rm c}$ for which in the phase diagram in Fig.1(a) a continuous PM/AF
transition takes place. For $U/t = 8, 12$, and 16 the value of $n_{\rm c}$
is (approximately) equal to 0.8 in each case. \\

\end{document}